# Design and Characterization of Crossbar architecture Velostat-based Flexible Writing Pad

Thesis submitted in partial
fulfillment of the requirements for
the degree of

*Master of Science in* **Electronics and Communication Engineering** *by Research*

by

Mohee Datta
Gupta 2018112005
`mohee.datta@research.iiit.ac.in`

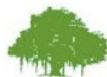

International Institute of Information
Technology Hyderabad - 500 032, INDIA
June 2023

International Institute of Information Technology

Hyderabad, India

## **CERTIFICATE**

It is certified that the work contained in this thesis, titled **"Design and Characterization of Crossbar architecture Velostat-based Flexible Writing Pad"** by Mohee Datta Gupta, has been carried out under my supervision and is not submitted elsewhere for a degree.

\_\_\_\_\_\_\_\_\_\_\_\_\_\_  \_\_\_\_\_\_\_\_\_\_\_\_\_\_\_\_\_\_\_\_\_\_\_\_\_\_\_\_\_\_\_\_

Date  Adviser: Dr. Aftab M Hussain

To getting closer to the dreams I wake up to

# Acknowledgments

I begin my acknowledgments by expressing my regards to my advisor, Dr. Aftab Hussain for entrusting me with the projects that comprise this dissertation. His advice and guidance were invaluable for me to complete my research. Most of all, Aftab sir always gave me the freedom to choose what I wanted to do and explore new things. I strongly believe that this freedom helped me shape my career as it stands now. I would also like to take this opportunity to thank my lab mates in the PATRIoT+FleCS lab without whom research might not have been fun. Special mention to L. Lakshmanan who was critical in helping me with experiments. I would not have been able to complete my projects in time had he not put in the hours to set me up for success. This thesis wouldn't have been possible without Anis Fatema, a co-author in almost all my papers. Thank you for all your support ma'am. I would also like to extend my gratitude to Animesh Sinha who helped me brainstorm and debug some critical parts of my project.

I am grateful to Deeksha Devendra for being my friend and mentor. She helped me to get through research. Her advice and words of reassurance were invaluable for my career and personal growth. It is said that we are the average of our closest friends. Thus I would like to thank (in no particular order) Srinath Nair, M. Sridhar, Arohi Srivastava, Yash Bhansali, Dolton Fernandes, and Shreya Malkurthi for their constant support and love which made college and research fun. I am also grateful to Trisha Chakravorty for being a good friend and helping me edit this thesis. Without her help, the thesis might have been littered with spelling and grammatical mistakes.

Finally, I would like to thank my parents for constantly motivating me to outperform myself and always strive for success.

# Abstract

Pressure sensors are popular in a large variety of industries. For some applications, it is critical for these sensors to come in a flexible form factor. With the development of new synthetic polymers and novel fabrication techniques, flexible pressure sensing arrays are more easily accessible and can serve a variety of applications. As part of this dissertation, we demonstrate one such application of the same by developing a low-cost flexible writing pad and doing crosstalk analysis on sensors with similar working principles.

Digital multimedia tools are becoming increasingly important as the world slowly shifts to an online mode. We present a low-cost, flexible writing pad that uses a 16×16 pressure sensing matrix based on the piezoresistive thin film of velostat. The writing area is 5 cm × 5 cm with an effective pixel area of 0.06 mm$^2$. A read-out circuit is designed to detect the change in resistance of the velostat pixel using a voltage divider. A microprocessor raster scans through the sensor pixel matrix to obtain a data frame of 256 numbers. This data is processed using techniques like squaring and normalising (S&N), Gaussian blurring, and adaptive thresholding to generate a more readable output. The writing pad is able to resolve characters larger than 2 cm in length. The flexible writing pad produces legible output while flexed at a bending radius of up to 4 cm. Such flexibility promises to enhance the usability and portability of the writing pad significantly. We noticed that the raw data produced by the writing pad had a lot of crosstalk which we were subsequently able to resolve using the algorithms mentioned above. Such crosstalk has been reported in literature multiple times and is common, especially for sensors of the crossbar architecture.

Crosstalk, in a sensor matrix, is the unwanted signal obtained at a sensor pixel that is not directly related to the stimulus. This paper presents a novel approach towards quantifying the crosstalk characteristics of a sensor matrix. The method involves obtaining the normalized sum of all the neighboring pixel readings, weighted by their distance from the stimulated pixel. We have used this methodology to characterise the crosstalk for a 5×5 velostat-based flexible pressure sensing matrix that uses a crossbar electrode architecture. We characterized sensor matrices with three different pitch lengths: 3mm, 4mm, and 5mm. We observe the crosstalk values to lie between 0.032 to 0.17 across different weights and pitches, which indicates a low measure of crosstalk. We observed that the 5 mm pitch matrix has the least crosstalk, which was expected due to the larger spacing between the pixels. We also characterised the crosstalk for all 25 pixels of a 4 mm pitch matrix and found the mean to be 0.081±0.002, within the range from 0.0071 to 0.1656.

# Contents

# List of Figures

# List of Tables

*Chapter 1*

# Introduction

In the present world of intelligent sensing and human-computer interaction, there has been an enormous evolution in the research and development of flexible and wearable sensing systems. Many exciting research works are currently being carried out in brain-machine interfacing, bioengineering, and biohack- ing, for which we require soft, curvilinear, and flexible free-form electronic systems rather than rigid, rectangular, and brittle systems. Flexible and wearable sensing technologies are emerging numerously and are performing a myriad of physical and physiological measurements [78]. The expeditious development and implementation of such sensors in the last decade have exemplified the importance and potential benefit of this distinctive class of sensing platforms.

Wearable technologies are procuring huge provocation due to the great demand for wearable devices like activity trackers in the form of smartwatches and bracelets. The basic element in all these devices is sensors, and the flexibility of these sensors is the major challenge. The sensing systems that detect and monitor the environment and communicate the obtained physical information, such as force, humidity, pressure, strain, and motion, form a fundamental building block of advanced applications. These applications include but are not limited to smart textiles [65], soft robotics [17], consumer and portable electronics [55, 31, 40, 41] aerospace [36, 2], automobiles [51, 58], biomedical [6, 28, 8, 48], and
environmental monitoring [61, 10].

The indefinite applications of sensors in each and every field have resulted in an extraordinary escalation of research in this area. Most of the applications mentioned above rely on force or pressure sensors, considering we can use force and pressure as synonyms in this context, as pressure is the force applied over an area where we are assuming that the area is known. Pressure sensors are widespread as they find many applications in aerospace [37], healthcare [7], wearable electronics [35], robotics[46], and so on. Fields like wearable electronics and sensing in healthcare benefit if these sensors can be made flexible [23]. With the recent development of new materials and novel fabrication procedures, flexible pressure-sensing arrays are more readily available. These flexible devices can either use pressure sensors based on piezoelectricity [34], piezoresistivity [29], or a change of capacitance [45, 27].

In the current era, especially after the Covid-19 pandemic, the world is more inclined toward an online setting for many activities. Therefore, digital multimedia tools are becoming more relevant. These modern tools include interactive whiteboards, graphic tablets, audio/visual input/output devices, webcams, etc. These technological innovations not only improve the training and learning experience but also are more effective than their traditional counterparts [56]. In addition to these factors, graphic tablets are also actively being used in the medical and healthcare industry [24]. The modern digital writing pads/graphics tablets typically work on the principles of capacitive touch [74], resistive touch [72], and Electromagnetic Resonance (EMR) technology [70]. However, these writing pads are rigid, limiting their portability and usability. Flexible electronics is an emerging field involving the development of wearable technology, stretchable electronics, printable electronics, electronic skin, etc. [23, 38]. The development of such devices, especially that of a digital writing pad, has the potential to improve the experience and efficiency of the user.

## 1.1 Scope of Thesis

This thesis comprises of two main works. First, we present a low-cost flexible writing pad based on a pressure sensor matrix using a velostat thin film. The writing pad has 256 pressure sensing pixels covering a writing area of 25 cm$^2$. As per our knowledge, there has not been any demonstration of a resistive pressure sensor based flexible writing pad. We have applied enhancement techniques like squaring and normalising (S&N), Gaussian blurring, and adaptive thresholding to generate a more readable output. The flexibility of the writing pad can provide an improved writing experience to the user, similar to using paper. The information written on it can be reproduced easily in digital form and can be saved on a smart device, computer or on the cloud.

Second, we present a novel approach to quantify crosstalk in a sensor matrix that can be extended to any array and used across technologies. To the best of our knowledge, there have been no such attempts in the past. This proposed method takes into account the distance of the neighbouring pixel that is added to the crosstalk and penalizes pixels which are further away but have a significantly high sensor reading.

We characterised three flexible pressure sensing matrices based on a crossbar architecture using velostat. We also used different weights to analyse the trends with changing pressure values.

## 1.2 Thesis Layout

The thesis is organized as follows:

- **Chapter 2:** In this chapter, we discuss the prerequisites or background work that is necessary to understand the work presented in future chapters. We look into what are pressure sensors, the different types of pressure sensors, their working principles, etc. A bit more detailed discussion is done on velostat, which is our piezoresistive pressure sensor of choice. All the sensors developed as a part of this thesis are designed on the crossbar architecture. Thus there is a detailed discussion on how it works and what are some of the past research done with it.

- **Chapter 3:** In this chapter, we look into the development of the low-cost flexible writing pad that was mentioned earlier. The fabrication process and the post-processing algorithms are discussed. The details of the experiments conducted and the observations made are provided. There is also a discussion on how our device currently stands up to a market ready product, and the scope of extension of this work in the future.

- **Chapter 4:** In this chapter, we briefly look at what is crosstalk and what have been the past attempts to resolve it. Then the proposed algorithm to quantify crosstalk is discussed with an example. We discuss the experiments conducted and what we infer from them.

- **Chapter 5:** In this chapter, we conclude with a summary of the methods and results discussed in this thesis.

- **Appendix A:** Here we look into some additional results of crosstalk calculated using the algorithm proposed in Chapter 4. The data for these were collected as a part of another project which aims to investigate the reliability of velostat as a pressure sensor. The details of this project is beyond the scope of this thesis.

*Chapter 2*

# Crossbar Architecture with Velostat as Sensing Material

## 2.1   Introduction

This dissertation uses flexible pressure sensors developed by us in all the upcoming chapters. Therefore in this chapter, we will first discuss what a pressure sensor is, what the different types of pressure sensors are, and why we prefer one over the other for a certain application. We will get into more detail on velostat, the pressure sensor we used to fabricate our sensors. These sensors are designed in the crossbar architecture which will be discussed in the following sections of this chapter. Finally, we will look at some of the existing literature that already exists on pressure sensors in crossbar architecture.

## 2.2   Pressure Sensors

A sensor is a device that translates any physical measurement into a signal, in most cases into an electrical signal. Therefore a pressure sensor is an instrument that generates a signal as a function of the pressure applied on it. Generally, pressure sensors are based on piezoelectric [34], piezoresistive [29], or capacitive sensors [27, 39, 49, 47, 45].

Let us discuss these three types of pressure sensors in more detail so that we can understand why we chose to proceed with one over the others.

- **Piezoelectric Sensors:** Piezoelectricity is the ability to generate electric charge in response to applied mechanical stress. Piezoelectric sensors are characterized by having a transduction element made of a piezoelectric material. Commonly used piezoelectric materials are lead zirconium titanite (PZT) ceramic, aluminum nitride (AlN), [1] etc. Piezoelectric pressure sensors usually have a sensing element that transmits the fluid pressure to the said transduction

element. In a well

designed sensor the effective area of the diaphragm is constant and therefore, the force transmitted to the transduction element is directly proportional to the acting pressure. This force is again converted into a proportional electric charge [19]. The limitation of using a piezoelectric sensor is the complex structure of the electronic interface that needs to be designed, as we require a charge amplifier to convert the high impedance charge output to a voltage signal [42].

- **Piezoresistive Sensors:** Piezoresistivity is a electromechanical effect that is characterized by a change in electrical resistivity in a material with applied mechanical change. This change is commonly a reversible microstructural change, such as a change in the degree of electrical continuity in the material. The sensing material in a piezoresistive pressure sensor is a diaphragm formed on a silicon substrate, which bends with applied pressure. A deformation occurs in the crystal lattice of the diaphragm because of that bending. This deformation causes a change in the band structure of the piezoresistors that are placed on the diaphragm, leading to a change in the resistivity of the material. This change can be an increase or a decrease according to the orientation of the resistors. Since a piezoresistive material must be electrically conductive, metals and carbons are the most widely used constituents of piezoresistive materials, which include composite materials with conductive constituents. Among the composite materials, polymer–matrix composites are dominant, due to their low fabrication cost [9].

- **Capacitive Sensors:** Capacitive pressure sensors measure the change in the electrical capacitance that is caused by the movement of a diaphragm on application of pressure. The basic working principle is as follows, a capacitor consists of two parallel conducting plates separated by a small distance. One of the plates acts as the diaphragm that is displaced by the pressure, changing the capacitance of the circuit. They can be operated over a wide range of temperature, have good repeatability of measurements, and can measure a wide range of pressures from vacuum to high pressures [15]. They are characterized in normal, transition, touch and saturation modes [44]. One of the main disadvantages of capacitive sensors is the non-linearity of the output, the conditioning circuit, and the complex computations needed [20, 53] .

Fig. 2.1 by Xu *et al.* summarizes the above accurately. While piezoelectric and capacitive pressure sensors require complex circuit design and computations to achieve their goal, the read out circuitry for piezoresistive sensors is relatively simple [13, 15]. Moreover, such sensors are often low-cost, more durable, and provide a higher resolution. Therefore, it is chosen as the pressure sensing material in this dissertation.

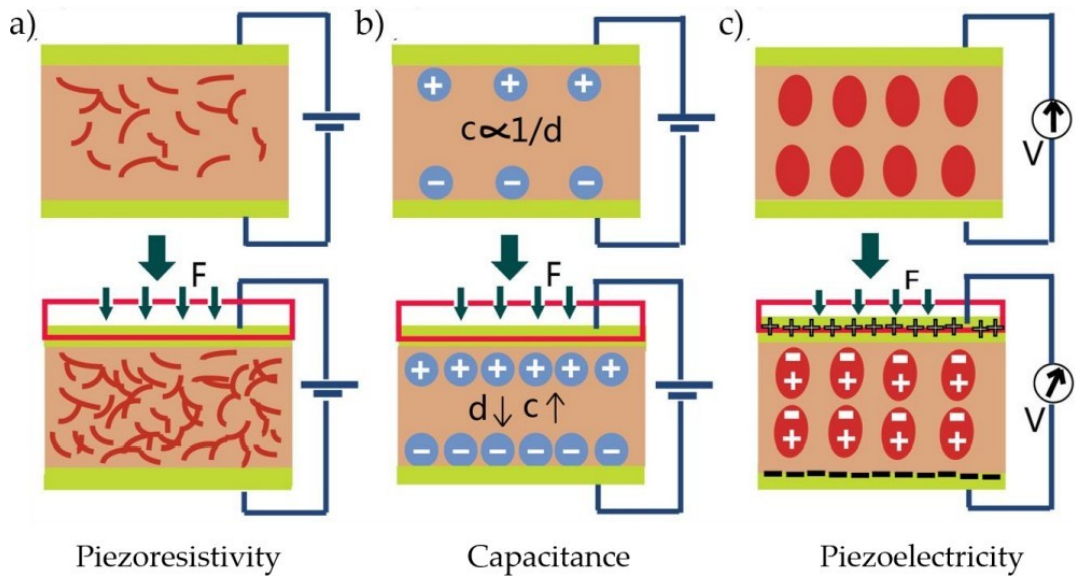

Figure 2.1: Schematic illustration of three common transduction mechanisms and representative devices: (a) piezoresistivity; (b) capacitance; and (c) piezoelectricity [75]

### 2.2.1 Velostat

Velostat is a part of a group of piezoresisitive sensors known as force sensing resistor (FSR). Force Sensing Resistor (FSR) is an analog sensor that function to change the compression force to change the resistance, when given the force on the FSR, the resistance will decrease. FSRs are made of a conductive polymer that changes resistance with force; applying force causes conductive particles to touch, increasing the current through the sensors [68]. Yuan *et al.* provide SEM (Scanning Electron Microscope) images of velostat with 5000x magnification. We can clearly see the physical changes that the material undergoes when it is under stress. In Fig. 2.2, the white spots are the carbon particles while the black spots are the gap in polymer clusters. While the average gap between the polymer clusters is $1\mu$m when under no pressure, it becomes $0.6\mu$m when placed under an external pressure [79].

Velostat has been used as a pressure sensing material in various research applications including finger gesture recognition [25], smart chairs [71], foot print pressure system [52], in-socket pressure sensing [21], wearable sensors [59], real time tracking system [57], etc. Velostat is predominantly

selected

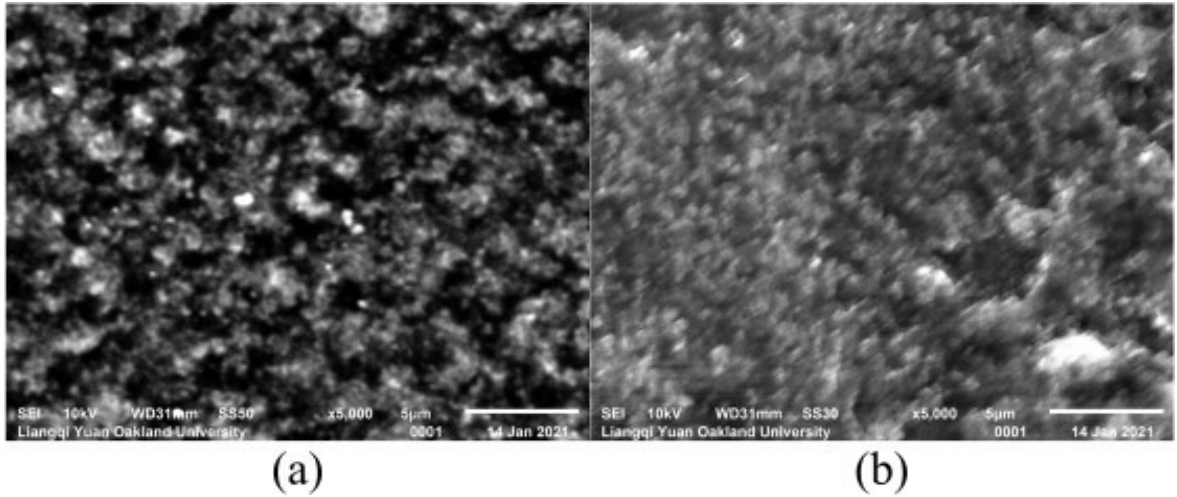

Figure 2.2: SEM image of Velostat material at 5000 times magnification, (a) without pressure and (b) with pressure. [79]

because it is inexpensive when compared to other piezoresistive materials such as graphene-based polymeric composites, porous graphene, and carbonized melamine which are difficult to fabricate [16]. Del Prete *et al.* studied the metrological properties of velostat and report good performances for response function, calibration, repeatability, sensitivity, time drift, hysteresis, and dynamic response[11]. However, all the above works report observing "phantom" readings that were not quantifiable in areas where no direct pressure was applied. This is the typical crosstalk effect in tactile sensors. We will discuss it in more detail in the following sections.

## 2.3 Crossbar Architecture

In this section, we look into the details of the crossbar architecture which will be used to fabricate sensors in future chapters. The crossbar architecture typically consists of three layers. The top and bottom layer contain the electrodes, with a pressure sensing layer sandwiched in the between the two. The top and bottom electrodes are orthogonal to each other in orientation, thus forming a mesh-like structure. The middle layer being a piezoresistive sensor, in our case velostat, will act as an insulator until an external pressure is applied. One of the layers, let's say the top layer will be set at a constant DC voltage while the other layer, the bottom layer, will be connected to a microcontroller which will read any voltage that is applied to the bottom layer. When an external pressure is applied on the mat the velostat's

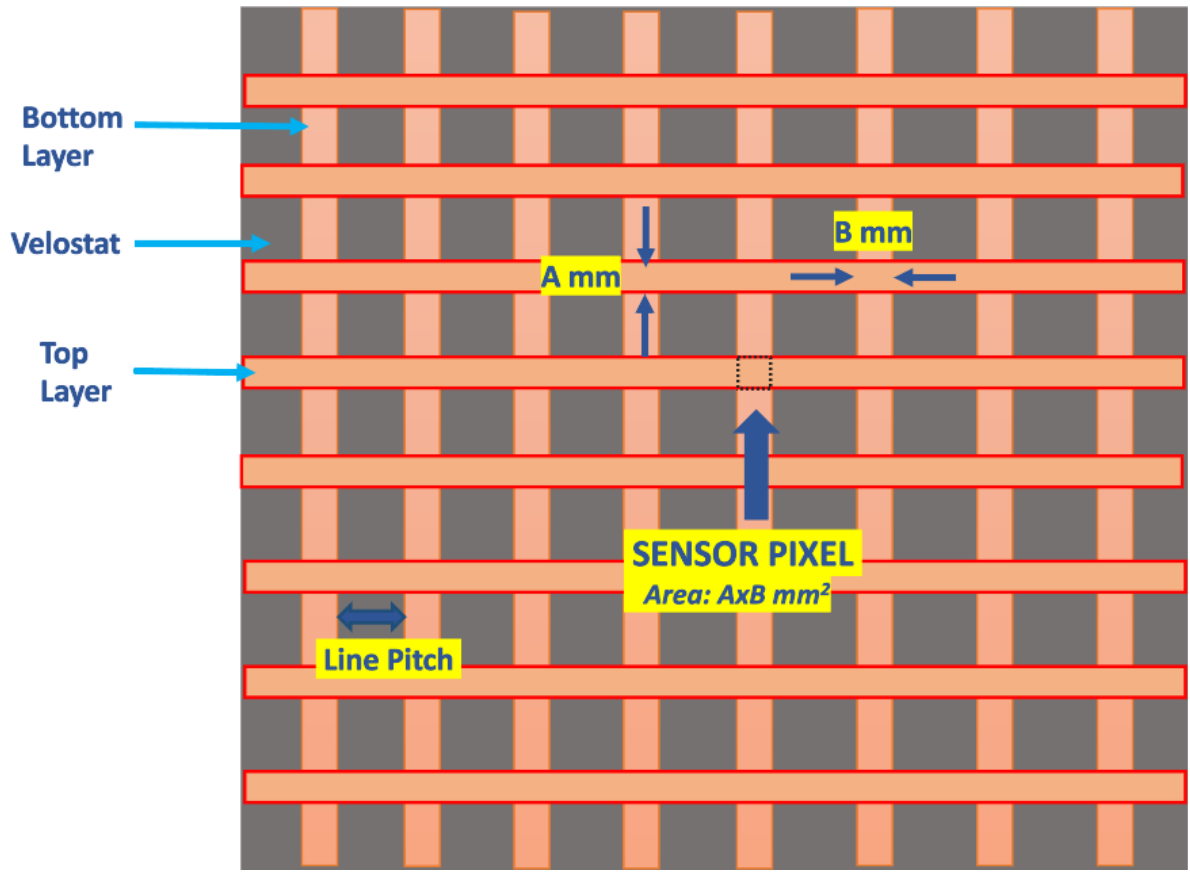

Figure 2.3: Schematic illustration of the top view of the sensor highlighting the sensor pixel formed by the overlap of top and bottom layer electrodes

resistance will decrease and a current will flow from the top layer to the bottom. We can observe that if there are $n$ lines of electrodes on the top and bottom layer each, there will be $n \times n$ overlaps across the net sensing area of the mat. Each such overlap will act as a sensor pixel. The sensor pixel area will depend on the width of the electrodes in the top and bottom layer. The line pitch or the distance between the electrodes will decide how far these pixels will be from one another. The closer they are, the higher the resolution of the sensor mat. The above sensor design is illustrated in Fig. 2.3.

### 2.3.1 Related Works

Several implementations of the sensor design mentioned above have been reported in literature. Sundholm *et al.* present a flexible pressure sensing mat with a sensing area of 80 cm × 80 cm that is used to detect gym exercises [67]. They used 80 electrodes with a pitch of 1cm for each of the two layers with a conductive textile as a sensing element. From the pressure readings collected by this mat, they were able to classify between 10 different gym exercises with an accuracy just less than 90%. Suprapto *et al.* created a 16×16 sensor matrix in the crossbar architecture for foot pressure measurements [68]. They used velostat as the piezoresistive material. The sensor matrix they developed was compared against the gold standard of human plantar pressure detection and reported satisfactory results. Gala *et al.* developed a system that can detect user presence in a chair using velostat based pressure sensors that send real-time pressure data to a website which can then be used to set customizable alerts. These sensors can not only detect the presence of a user but can provide details on the anatomical well-being of the user, like detecting scoliosis [18]. Yuan *et al.* developed a 27×27 velostat based pressure sensing array that can be used for object recognition. From the pressure images collected from the mat they used the neural network ResNet-PI to classify 10 objects, and reported an accuracy of 0.9854 for the same [79]. Our lab group has done several projects on the same. In [16] a 4×4 low-cost, flexible pressure sensing matrix was developed for activity monitoring and tracking recovery in stroke patients. Anis *et al.* further tested the mechanical reliability of these velostat based sensors in [14]. They report an observed deviation in output voltage was 0.95% for 15 mm, 0.95% for 20 mm, 0.97% for 25 mm, and 2.2% for 30 mm bending radii, for 150 bending cycles, with respect to the flat position. They also proposed a two parameter calibration model which can be used to minimize these deviations further. Ivin *et al.* developed a pressure sensing suit which functions on similar principles. This suit can detect incorrect postures for specific exercise techniques in real time [33]. They propose an algorithm which can classify each repetition (rep) of exercises as a good rep and bad rep with an accuracy of 95.5% across three subjects.

*Chapter 3*

# Development of Flexible Writing Pad

## 3.1 Introduction

In this chapter, we will look into the details that went into the development of a low-cost piezoresistive flexible writing pad. We start by taking the reader through the existing literature on a flexible writing pad. This will help us better understand the gap in research and hence the motivation for this work. Further, we discuss what the various hardware components used were and why. In the upcoming sections we present the different experiments conducted and the metrics that we used to present our results. In section 3.3, we explain the results we observed and how they compare to a market ready version of a similar product.

### 3.1.1 Related Work

The existing literature for a flexible writing pad includes work by Mochizuki *et al.* [50], where flexible transparent electrodes have been fabricated using PEDOT:PSS and resistive touch panels. The electrodes were fabricated using line patterning of conductive inks. They reported that the drawing on the resistive touch screen panel was successfully displayed on the PC screen with good in-plane uniformity and linearity. Kim *et al.* demonstrate a dynamic interactive sensor capable of arbitrary visible pattern generation on a thermochromic panel with high resolution [30]. They used this technology to demonstrate a potential prototype of a thermochromic writing board. They could alter thickness of lines by varying the local pressure applied, thus realizing a dynamic pressure mapping with multilevel sensing capability. The two studies mentioned above are the closest to our research but they are not trying to build a similar product. Moreover they rely on organic polymers as their substrates. This not only makes fabrication of such a product expensive but also hard to scale. There are some other works which claim that a flexible writing pad can be one of the applications of their innovation. For example, Nair *et al.* fabricated a transparent capacitive touch pad in the form of a 2×2 matrix [54]. A silver nanowire-based ink was used to print the electrodes with PDMS as the dielectric. The touch pad thus created has a high degree of flexibility. They claim that the formulated nanowire ink can be extended for other flexible and stretchable transparent sensing applications. Shi *et al.* have used a flexible wearable triboelectric patch with four electrodes capable of detecting unique patterns drawn on them by distinguishing the unique patterns in voltages generated [62]. With this patch they can achieve position sensing with clear differentiations even under different types of operations including both tapping and sliding interactions. Pu *et al.* also used a flexible triboelectric 3D touchpad which can be used for various Human-Machine Interface (HMI) operations [60]. Their design consisted of a multi-channel positioning layer and a single-channel pressure sensing layer. Kim *et al.* used a novel input mechanism based on finger writing on a flexible textured pad [32]. Their design uses patterned vibrations to recognise different gestures like tapping, rubbing, flicking, etc. Reflex Display Technology and its various potential applications have been discussed in [26]. They present a flexible touch-sensitive writing tablet utilizing a reflective bistable cholesteric liquid crystal laminated between two conductive-polymer coated polyethyleneterephthalate substrates. The device features a selective pressure response suitable for high-resolution lines to be drawn into the screen. The report that the said device is inexpensive, conformable, and laser-cut to any desired shape. '

### 3.1.2 Bending Radius

In this study, we created a flexible writing pad. For characterizing the flexibility of our device, we need a parameter to quantify the flexibility of an object. The most popular method is to use the minimum bending radius of an object. In this section we will explore what is bending radius, how to calculate it, and why it can be used as a parameter to quantify flexibility. Further details may be found in [23].

The perceived flexibility, or lack thereof, of an object around a particular axis of bending is called its flexural rigidity. In other words, this is the resistance offered by a body on being subjected to a bending force. Thus, the lower the flexural rigidity, the higher the perceived bending. To derive a mathematical relationship of flexural rigidity with material properties of an object we must focus on the concept of curvature. We can calculate the curvature of any point on a 2D curve by finding the radius of the circle approximating the curve at that point. This can be better understood from Fig. 3.1 [23], where for a

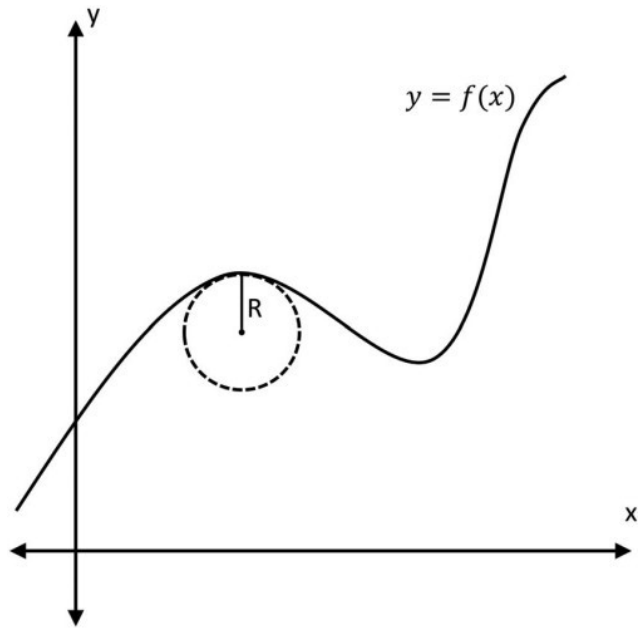

Figure 3.1: Illustration of Bending Radius [23].

function $y = f(x)$ the curvature $\kappa$ at point $P$ is given by:

$$\kappa = \frac{|y''|}{(1 + (y')^2)^{3/2}}$$

Now we come to our parameter of interest, the bending radius. The bending radius $R$ is simply the inverse of the calculated curvature of the structure.

$$R = 1/\kappa$$

Thus a lower bending radius implies that a structure can be subjected to a larger curvature while bending, i.e., makes it more flexible. Therefore the lower the minimum bending radius of an object, the more flexible it is.

## 3.2 Materials And Methods

### 3.2.1 Writing Pad Design

The flexible writing pad has three critical components in its design:-

- **PC:** The device is powered by the PC and all the post-processing algorithms are run on the PC. We will discuss the said algorithms in more detail in the upcoming sections.

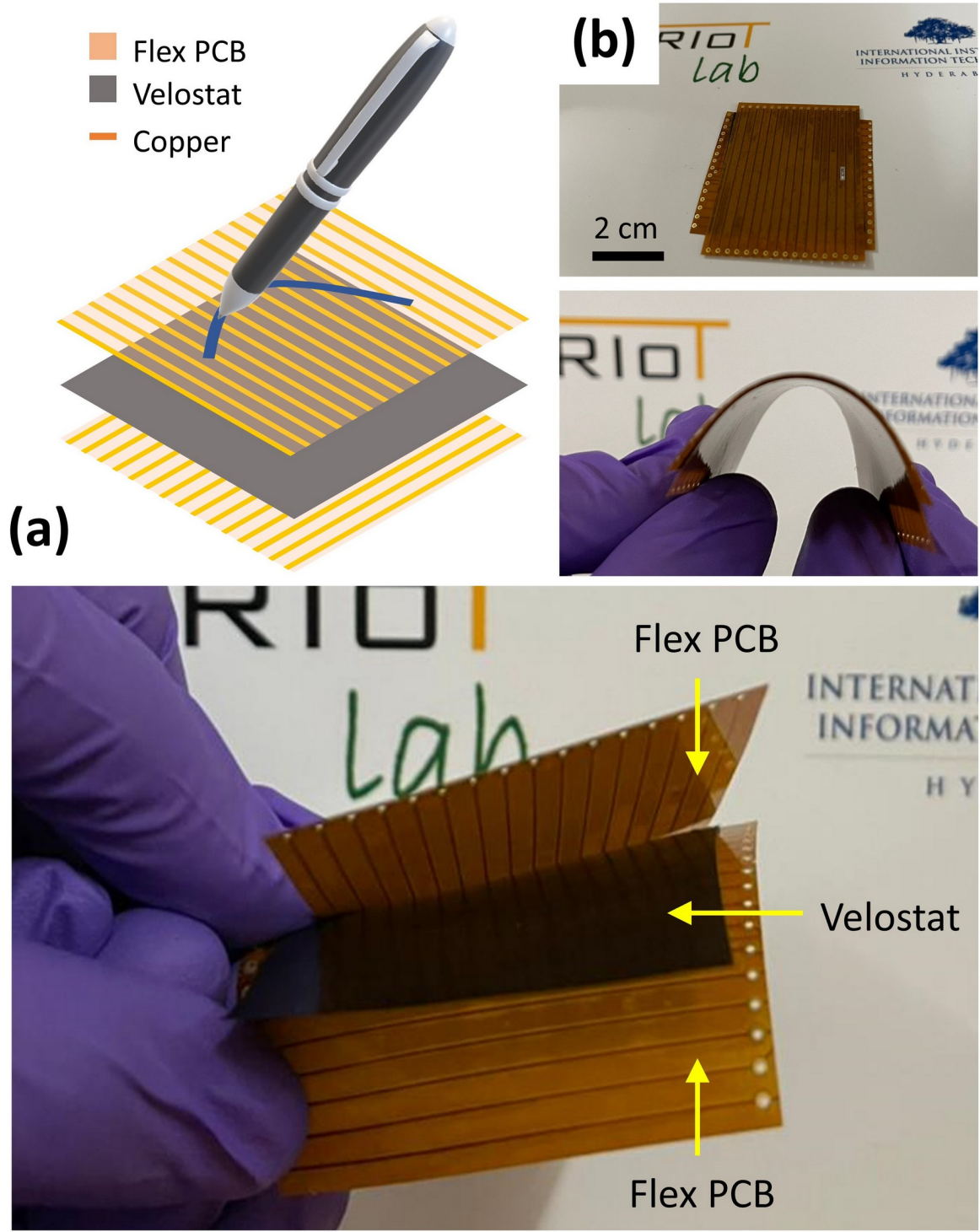

Figure 3.2: (a) Schematic illustration of the sensor system designed by sandwiching a sheet of velostat between two flexible PCBs. (b) Photographs of fabricated sensor displaying its flexibility, internal layers and crossbar architecture.

- **The readout circuitry:** This consists of the microcontroller, and the MUX and DEMUX combi- nation. The microcontroller in our case is the Arduino Nano 33 IoT. The Arduino Nano 33 IoT combines the Arduino Nano form factor with an easy point of entry to basic IoT and pico-network applications. We use 74HC4067DB as our MUX. The multiplexer, shortened to "MUX" or "MPX", is a combinational logic circuit designed to switch one of several input lines through to a single common output line by the application of a control signal.

- **Pressure Sensor:** This consists of the flexible PCBs and velostat. The two flexible PCBs used are made of a thin polyimide substrate. We print copper lines on it as electrodes to form the cross-bar architecture as discussed in Ch2. These two PCBs sandwich our piezoresistive material, velostat, between them to form our pressure sensor.

The flexible writing pad is a 16×16 array of piezoresistive pressure sensor pixels. It comprises of three layers as shown in Fig. 3.2a. The top and bottom layers are flexible PCBs made of polyimide substrate. We have printed copper lines on the PCBs in horizontal and vertical alignment such that they form a crossbar architecture such that each cross over point is a sensor pixel. The middle layer is a carbon-impregnated piezoresistive material named velostat. It is a low-cost, lightweight and flexible material making it an ideal choice for designing a flexible writing pad. The thickness of the PCBs are 180 $\mu$m each and that of the velostat is 106±2 $\mu$m (measured using Mitutotyo micrometer). The total thickness of the writing pad is approximately 460 $\mu$m which makes it highly flexible (Fig. 1b). The sensing area of the writing pad is 5 cm × 5 cm. Each copper line is 0.254 mm thick and is separated by a pitch of 3 mm. Hence, each sensor pixel is of area 0.254 mm × 0.254 mm.

The working principle of the writing pad has already been discussed in detail in Chapter 2. Just to quickly summarize, when a pressure is applied on velostat, its resistance decreases. This resistance is measured using a potential divider circuit formed using a fixed bias resistor. The bias resistor was selected as 1 k$\Omega$ to get high sensitivity [15]. Each pixel is selected by raster scanning through the matrix using a demultiplexer for voltage application and a multiplexer for read out (Fig. 3.3).

### 3.2.2 Algorithm

To generate the required output image of the pen strokes done on the mat, we performed a set of post processing operations on the data received by the microcontroller. First, the received data was converted into a 16×16 matrix, corresponding to the sensor array of the mat, for proper visualisation. We received real-time pressure data from the mat in frames of 0.1 sec each (after completion of one raster cycle). These frames were stacked together in an array. After receiving a fixed set of frames, say $n$, we added together all the frames to get a 16×16 sum frame. This is referred to, in this work, as the raw data frame. The value of $n$ can be determined by balancing a trade-off between the refresh rate of the writing pad and the signal-to-noise ratio (SNR). Higher value of $n$ gives a better SNR, while reducing the refresh rate. For all our experiments, we used $n$ as 100 and therefore with a frame rate of 0.1 sec we got a refresh rate of 10s. We found that it took on an average of 0.288s to apply post-processing operations, after the summing of the data frames. Keeping in mind that while writing some pressure would not have been applied to the area where the tip of the stylus touches the mat, we performed a squaring and normalising (S&N) operation on the raw data to give preference to the position with higher pressure data. This also eliminated most of the noise that we get as a result of crosstalk from neighbouring sensor pixels. After this, we applied a Gaussian blur filter on the image to further eliminate noise and get a smoother image. This was followed by adaptive thresholding to get a monochrome image. The threshold was calculated by determining the average pixel value of the smoothed image. Fig. 3.3 presents an overview of the above algorithm with sample outputs at each stage.

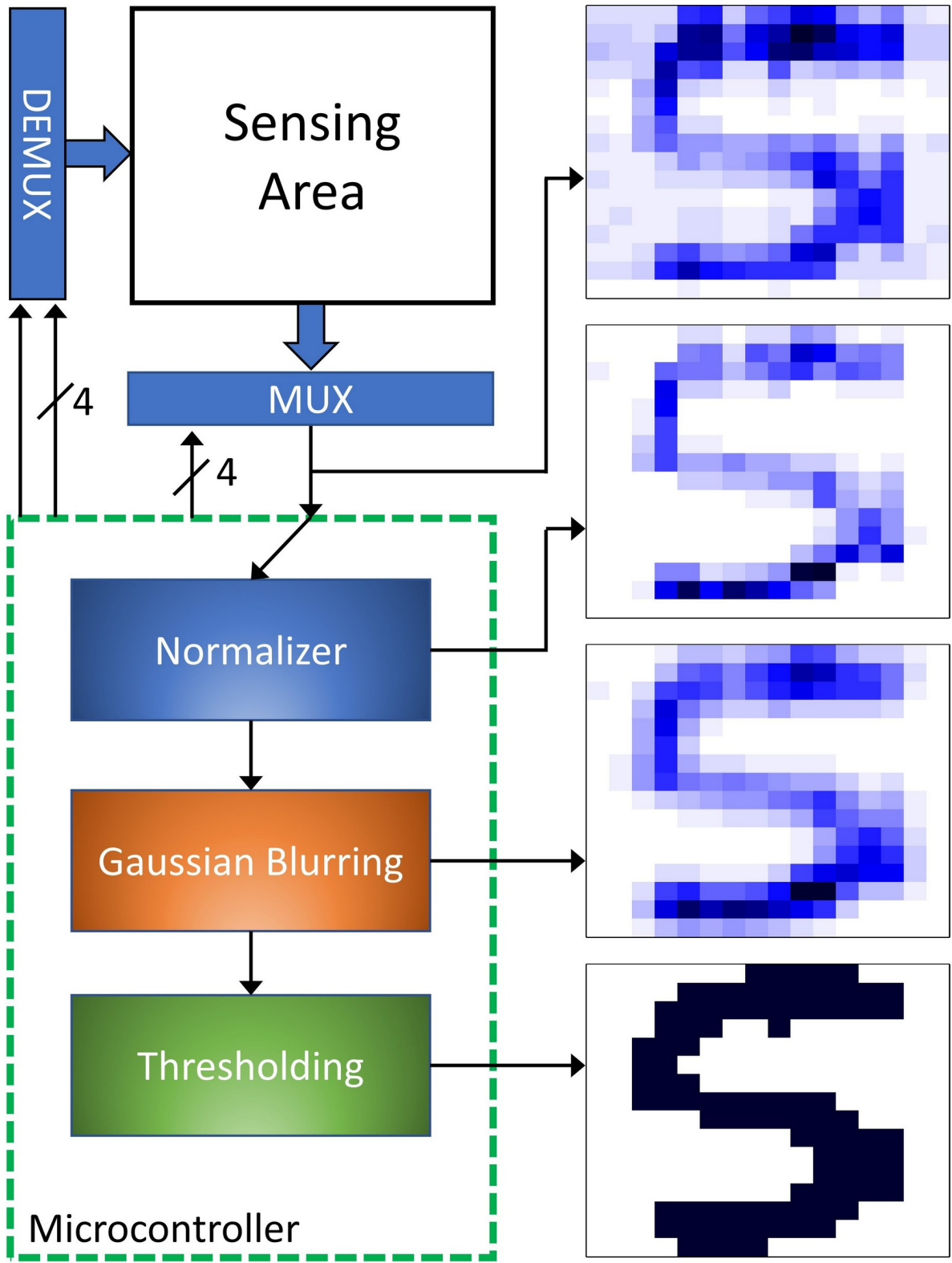

Figure 3.3: Overview of the data flow for the algorithm with sample outputs generated at each stage.

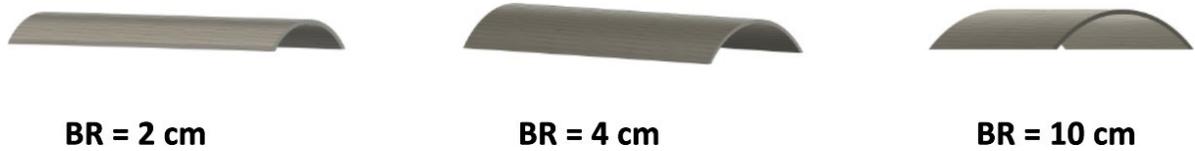

Figure 3.4: 3D models of half-cylinders with different radii

### 3.2.3 Experiments

To test the functionality of our writing pad we tested it for multiple probable use cases. We tried to recreate different letters of the English alphabet. We acknowledge that resolving smaller characters will be harder for our device and thus tested it by writing the same characters in decreasing size. As we claim that our device is flexible, we had to write on it while it was subjected to some sort of

bending stress and

observe how it affected the resolution of the characters. We 3D printed half-cylinders of various bending radii, approximately 2cm, 3cm, and 4cm, (as shown in Fig. 3.4) placed our device on the same and wrote on it. Then these results were compared to the ones while the device was kept on a flat surface. The device was also tested to see if it can resolve numbers of varying sizes correctly. All the observations made in the experiments discussed above have been presented in the following section.

## 3.3 Results and Discussion

As expected, the raw data shows a lot of crosstalk noise from the neighbouring sensors. This is mainly because of the cross-bar architecture of the electrodes and the use of a continuous film of velostat. However, we could achieve a good isolation of our region of interest by using the S&N operation. This particular operation was chosen over other higher order polynomial functions because they eliminated too much of the raw data and recovering operations on it were computationally expensive and did not yield good outputs. For Gaussian smoothening, a $\sigma$ of 0.6 yielded the best results and has been used for all experiments. Because our writing pad is flexible, we recreated the results while it is flexed. For this, cylindrical structures of bending radii 4 cm and 10 cm were 3D printed. Fig. 3.5 shows the results obtained when the writing pad was placed on these structures. We see that for 10 cm bending radius, the image is similar to that of the original. However, for bending radius of 4 cm, we experienced noise, particularly in the bottom part of the image. This can be explained by the stress/pressure put on the sensors due to bending of the pad.

The writing pad was also used to write different letters and numbers of various sizes. As shown in Fig. 3.6, we observed that the resolution of the captured image decreases as the size of the letters is reduced. The sizes of the objects shown in Fig. 3.6a are approximately 4 cm, 3 cm, and 2 cm, from left to right. This was expected because the smaller the object gets, the distance between two neighbouring strokes while writing it will decrease, leading to increased cross-talk noise. We also noted that for smaller alphabets using softmax instead of S&N yielded better results. However, all outputs presented in the paper have been generated using S&N for consistency. We observed that characters greater than 2 cm were easily legible, whereas for smaller characters, the SNR was very low.

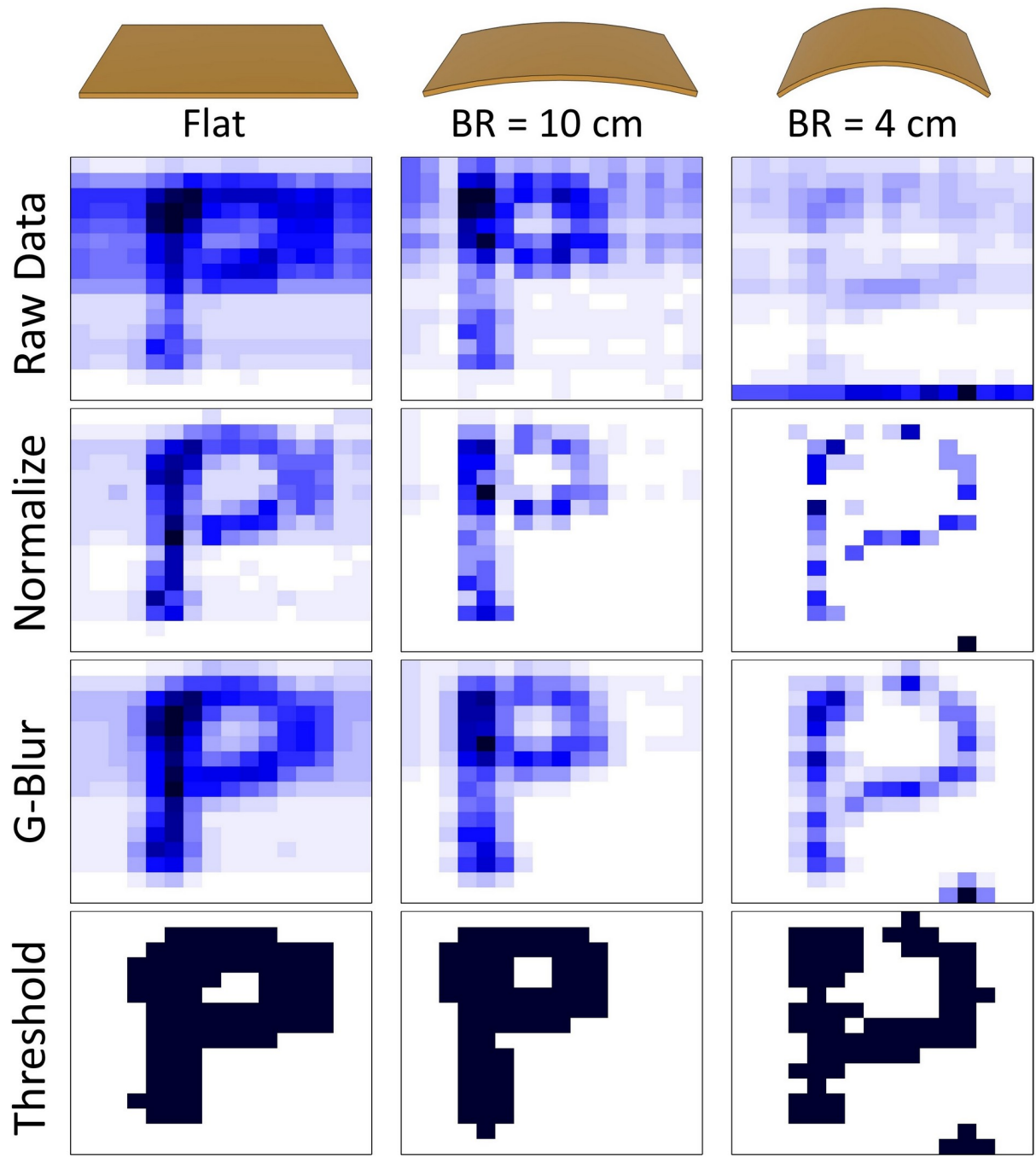

Figure 3.5: Results of experiments performed to verify the flexibility of the writing pad subjected to different degrees of bending.

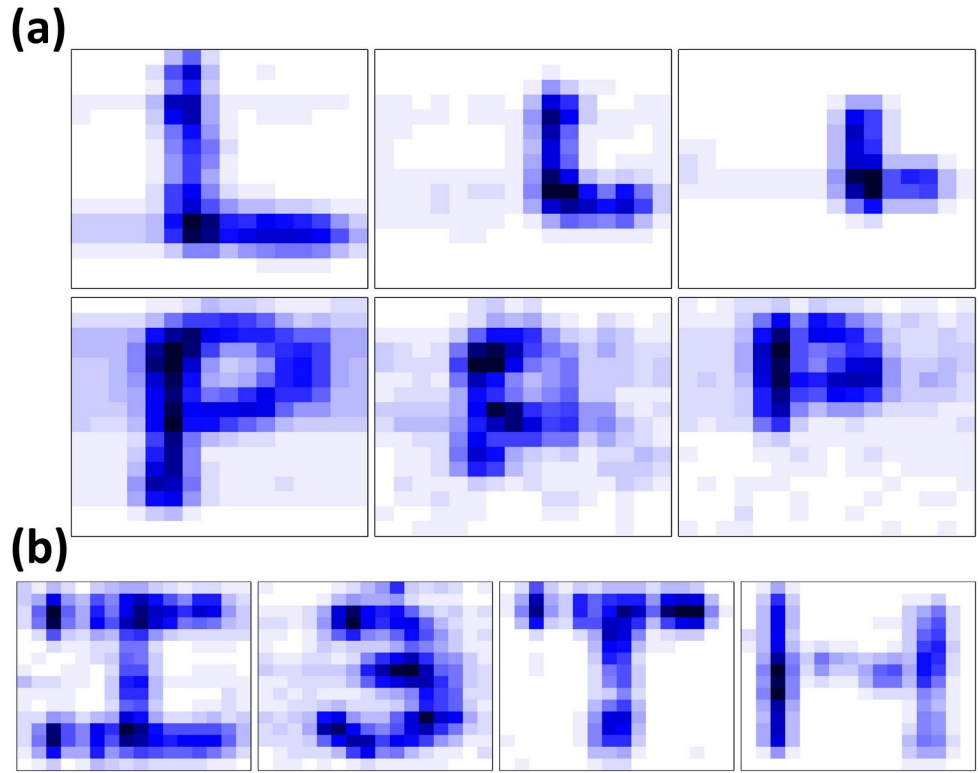

Figure 3.6: (a) Outputs comparing resolution of written objects of varying size. (b) An example text written on the pad representing I3TH. These outputs are after the Gaussian blurring step.

### 3.3.1 Future Scope

The device produced as a part of this project was a proof of concept and is not market ready yet. However, we feel that since this product has a chance to develop as a consumer device a comparative analysis with an existing market product would be interesting. We compare our device with an entry level digital writing pad that is commercially available. When comparing these results with the rigid digital writing pads currently available in the market we find that they provide a refresh rate of 20ms compared to our 10s. These boards are priced around $80-$100 and provide a sensor area of 15 cm × 15 cm. We fabricated ours under $20 with a sensing area of 5 cm × 5 cm, with our major expense being the microcontroller. Our microcontroller, as is, costs $15 which can be brought down significantly by developing a custom PCB and printing at scale. We can easily scale our device to have a higher sensing area while using this same PCB while keeping the overall cost similar. This is because the pressure sensing material, i.e., the velostat is inexpensive. This cost can be brought down even further by developing in bulk. However, no matter how cheaply we can produce the said device, to make this device market ready the refresh rate has to be improved. This can be achieved with a combination of a faster raster cycle, faster MUXes, and more optimised post-processing algorithms. The above mentioned commercial mats can detect upto 2048 unique pressure levels and have a resolution of 5000 LPI. "LPI" stands for lines per square inch, it is a measure of how many unique lines the tablet can measure/detect when drawn within an inch of its sensing area. We are yet to characterize our device for these specific parameters as they are not trivial, these numbers are estimates given as an output of various software approximating algorithms. We are working on these characterisations and as part of future projects in our lab group.

*Chapter 4*

# Characterisation and Quantification of Crosstalk

## 4.1 Introduction

In Chapter 3, we saw that there was a lot of crosstalk in the raw data frame which was later eliminated using the post-processing algorithms. Crosstalk like this is an acknowledged problem in sensors with crossbar architecture. In the upcoming sections of this chapter we will look into what is crosstalk, what are the causes of it, and how people have dealt with it in the other studies. Then we propose a novel method of quantifying and characterizing crosstalk in a velostat-based flexible pressure sensing matrix.

### 4.1.1 Related Works

Crosstalk is the phantom reading detected by a sensor due to the activation of neighbouring sensors. It is an accepted problem specially in context to the sensor designed in a crossbar architecture. It can be classified as mechanical and electrical. Due to limited space on the sensor array, each sensor pixel is placed close to each other and are thus mechanically coupled. This implies that each sensor pixel reading is not only dependent on the force applied directly on it but also relies on the conditions of its neighboring pixels [77]. To make the fabrication easy and scalable a single film of sensing material is used which leads to further force diffusion under pressure [5]. The electrical nature of crosstalk can be further classified into two categories [79]. The first is related to the surface resistance of the velostat which allows currents to flow through neighboring pixels. The second is related to the principle that unpressed velostat does not have infinite resistance. This leads to a loss of accuracy [69, 3, 12]. Attempts to reduce crosstalk involve placing diodes in series with each cell to avoid current flow [64, 76, 68], minimization of leakage current using pull-down resistors or setting the unused lines to a desired voltage to prevent any leakage altogether [73, 63]. Software attempts on resolving the crosstalk problem involve circuit analysis algorithms [43], as well as standard image processing algorithms aided with machine learning post-processing to generate a cleaner output [79, 66, 4, 22].

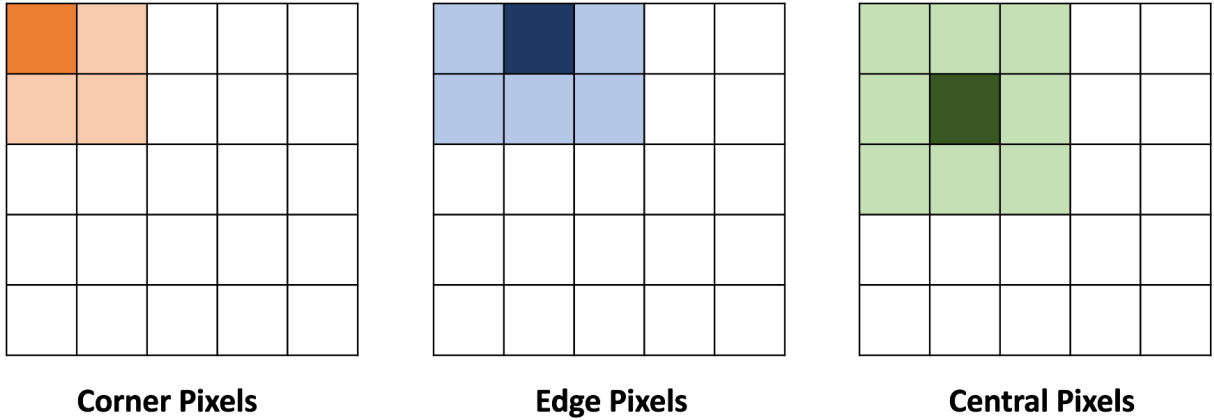

Figure 4.1: Illustration of the three possible neighbourhoods for a pixel on a matrix

### 4.1.2 Motivation

Even though there have been efforts towards minimizing crosstalk, to the best of our knowledge, there have not been any attempts to quantify it. To understand and improve any parameter the first step is to measure it and thus we believe that there should be a uniform scalable guide to empirically calculate and measure crosstalk. Therefore, in this work, we have developed a method to quantify the crosstalk in sensor readings. This proposed method takes into account the distance of the neighbouring pixel that is adding to the crosstalk and penalizes pixels which are further away but have a significantly high sensor reading.

## 4.2 Quantifying Crosstalk

The crosstalk measure presented in this paper is defined for every pixel in the matrix individually and can ideally involve all pixels in the matrix. For our purposes of characterising a $5 \times 5$ pressure sensing matrix, we only consider the pixels that are "adjacent" (edge and corner) to the pixel we are measuring. This gives us three types of neighbourhoods for the pixels in the matrix as shown in Fig. 4.1:

- **Corner pixels** with three neighbour pixels

- **Edge pixels** with five neighbour pixels

- **Central pixels** with eight neighbour pixels.

The measure is a function of the pressure values of these pixels as well as their distance from the pixel where the pressure is being applied. We weigh each of the neighbouring pixels according to their distance from the pixel where the pressure is being applied. If a single pixel is pressed, the pressure values should ideally be zero on all neighboring pixels, or at least should be inversely proportional to the distance between the pixels. So, if a pixel far away shows a high value, we can conclude that the crosstalk is high and the measure should indicate that by returning a high value. Thus, we have defined the crosstalk value at each pixel as the normalized sum of the pressure reading at neighbouring pixels weighted by their distance to the pixel under consideration. Hence, we get the following equation for crosstalk at a given pixel:

$$C = \frac{\sum_{i \in n - \{s\}} d_i p_i}{p_0 \sum_{i \in n - \{s\}} d_i} \quad (4.1)$$

where $p_0$ is the pressure value of the sensor we are applying pressure, and $p_i$ and $d_i$ are the pressure value and distance of the neighbouring pixels. As the neighbouring pixel readings will always be less than that of $p_0$, this parameter will be a dimensionless value between 0 and 1. We can use this equation for any size of the matrix and consider any size for the "neighbourhood" of the pixels, allowing the general structure of the measure to remain consistent.

We consider the reading we get from our pressure sensing mat for a weight of 500g on the bottom right corner pixel, which can be denoted as (4, 0) in Cartesian coordinates. The pressure matrix we obtain is shown below.

$$\begin{bmatrix} 0.00 & 0.00 & 0.00 & 0.00 & 0.00 \\ 0.00 & 0.00 & 0.00 & 0.00 & 0.00 \\ 0.00 & 0.00 & 0.00 & 0.00 & 0.00 \\ 0.00 & 0.00 & 0.00 & 0.00 & 0.15 \\ 0.03 & 0.04 & 0.04 & 0.04 & 1.94 \end{bmatrix}$$

In our case

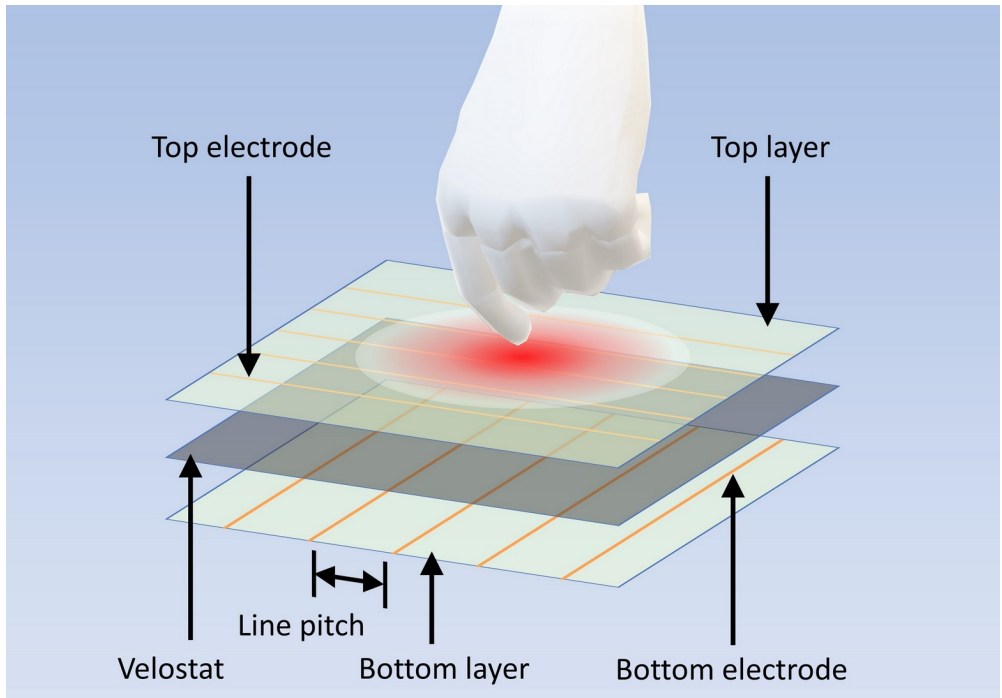

Figure 4.2: The construction of the velostat pressure sensing matrix with the illustration of the occurrence of cross talk.

We consider three pixels in this case. The Cartesian coordinates of the three neighbour pixels will be (4, 1), (3, 0) and (3, 1). Their Euclidean distances from the (0, 0) pixel will be 1, $\sqrt{2}$ and 1 respectively. Now, on substituting in 4.1, we get

$$C = \frac{0.15 \cdot 1 + 0.00 \cdot \sqrt{2} + 0.04 \cdot 1}{1.94 \cdot (2 + \sqrt{2})} = 0.02868 \tag{4.2}$$

This calculated number is what quantifies the amount of crosstalk at a certain pixel. As we can see, it is quite low, which is as expected.

## 4.3 Sensor Design

To test the crosstalk parameter, we designed flexible pressure sensing mats consisting of a 5×5 array of piezoresistive pressure sensing pixels (Fig. 4.2). The top and bottom layers were flexible PCBs made of polyimide substrate. We printed copper lines on the PCBs in horizontal and vertical alignment such that they form a crossbar structure with each cross-over point being a sensor pixel. A velostat sheet was used as the piezoresistive sensor. When pressure is applied the resistance of the velostat decreases. The resistance was measured using a potential divider circuit formed using a fixed bias resistor of 1 kΩ, to get high sensitivity [16]. As pressure is applied, the resistance of the pixel decreases and the voltage across it increases. This voltage was read through the analog input pin of the microcontroller. The thickness of the PCBs were 180 $\mu$m each and that of the velostat was 106±2 $\mu$m (measured using Mitutotyo micrometer). Thus, the total thickness of each mat was approximately 460 $\mu$m which makes it highly flexible. We decided to develop mats with three different levels of resolution. The three mats have a pitch, separating each copper line from the other, of 3 mm, 4 mm, and 5 mm (Fig. 4.3). Each copper line is 0.254 mm thick and hence, each sensor pixel is of area 0.254 mm × 0.254 mm.

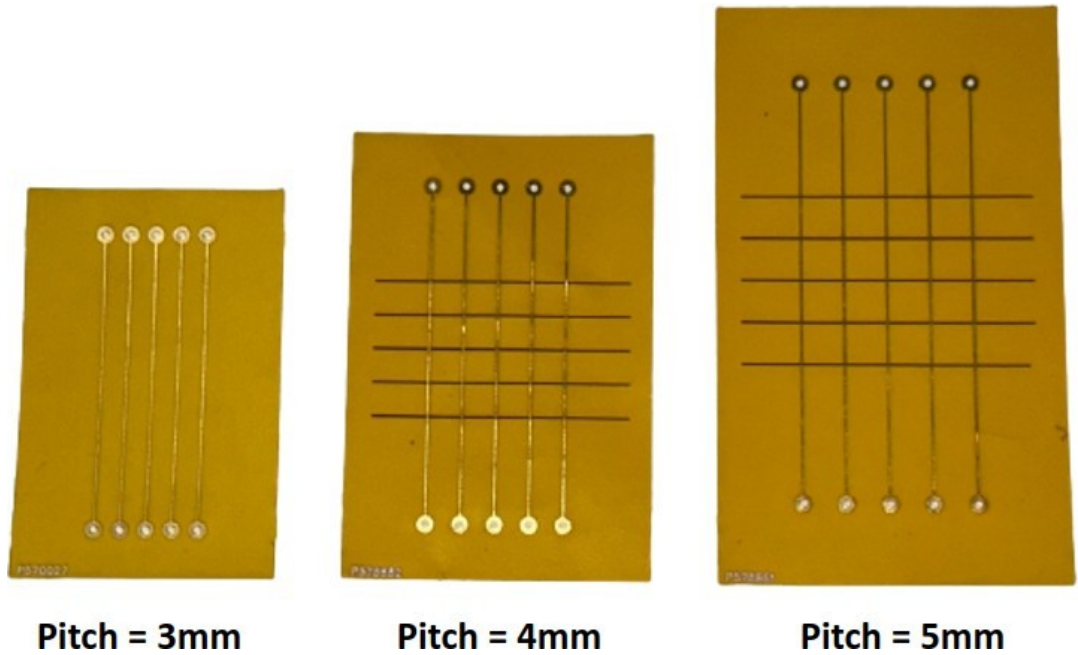

Figure 4.3: One layer of electrodes of the designed sensors with pitch of 3mm, 4mm, and 5mm respectively.

## 4.4 Experiment Setup

The crosstalk data has been collected using a specially constructed setup to obtain accurate readings. A force gauge was used to apply pressure onto a specific pixel, while the readout circuitry read data from all pixels. To maintain the amount of pressure applied onto the pixel for a reasonable amount of time to collect data, a stand structure was fabricated. The mat was connected to the readout circuitry and was placed onto an acrylic plate. The force gauge was inserted into the stand and positioned over the pixel under observation. A weight was then placed over the stand to apply a certain amount of pressure onto the pixel, and the force value was read out from the force gauge. The experimental setup is shown in the figure Fig. 4.4.

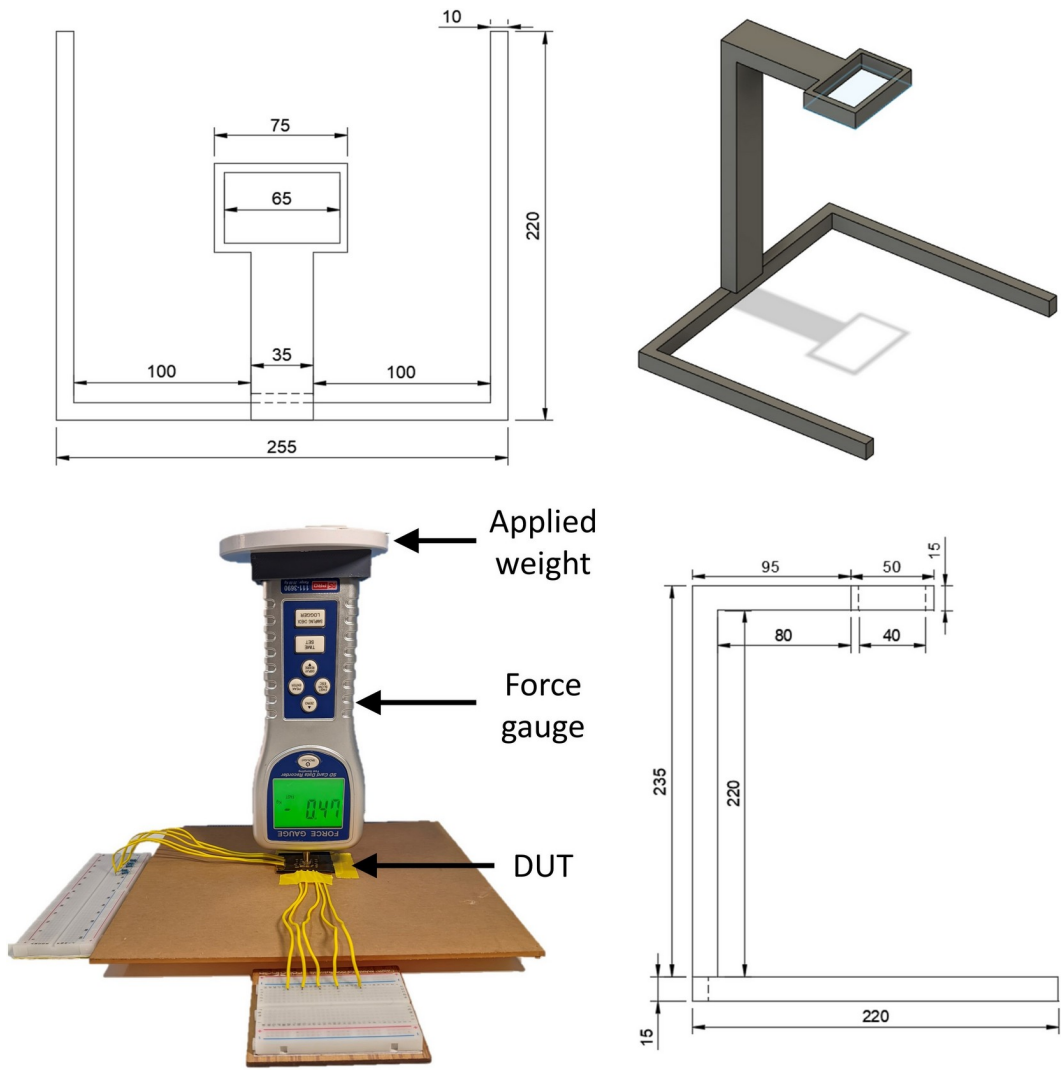

Figure 4.4: The design of the experimental setup to characterize the cross talk in device under test (DUT) for various applied weights and line pitches.

## 4.5  Results and Discussion

We performed two different sets of experiments in an attempt to characterize the crosstalk of the pressure sensing matrix. For the first experiment, we analyzed how the crosstalk varies as the applied pressure increases and how it is affected by the pitch of the sensors. A known force was applied on the centre pixel of the mat, the pressure data was recorded for all pixels and crosstalk value was calculated. This was repeated for different weights, 0.5 kg, 1 kg, 1.5 kg, for mats with line pitches of 3 mm, 4 mm and 5 mm. The results of this experiment are shown in Fig. 4.5. We can observe that the mat with a pitch of 5mm has the least crosstalk of the three mats which is expected as the further each neighbouring pixel is, the lesser its influence.

For the second experiment, we analysed if the position of a pixel on the mat influences the crosstalk. As discussed in the earlier sections, there will be essentially three different kinds of sensor pixels, the corners, edges and central pixels. We applied a known pressure on all 25 pixels of the mat individually and recorded the crosstalk values. We used the 4 mm pitch mat for this experiment and 0.5 kg as the known weight. Fig. 4.6 shows the crosstalk value recorded at each of the 25 positions. The mean crosstalk value was observed to be $0.081\pm0.002$, with minimum and maximum crosstalk of 0.0071 and 0.1656. This indicates the positional variance in crosstalk is not due to any inherent design or material property of the sensor but probably due to the experimental conditions and sensor mat fabrication process.

## 4.6  Future Scope

In this project, we observed the changes in crosstalk by varying the pitch of the sensor mat and the applied pressure. Another parameter that can influence the crosstalk is the thickness of each electrode as it is directly proportional to the area of each pixel. A similar comparison can be done by varying the electrode thickness. In Fig. 4.5, we observe that the intuitive trend is not followed. As we will

see in

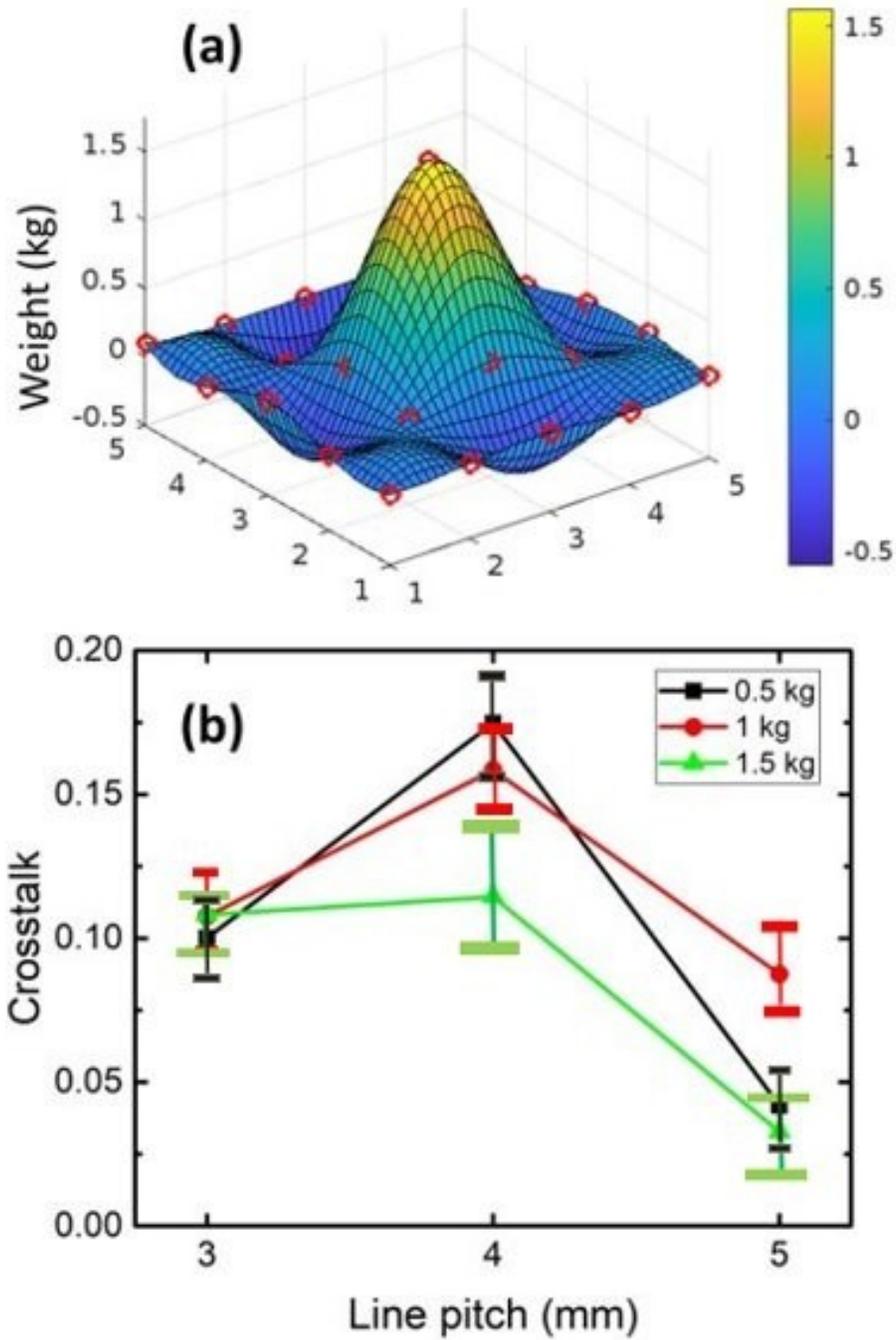

Figure 4.5: (a) The distribution of reported weight values for 1.5 kg weight applied at the center pixel for line pitch of 3 mm. Red dots indicate the sensor values at each pixel. (b) Crosstalk values calculated for the center pixel for various line pitches and applied weights.

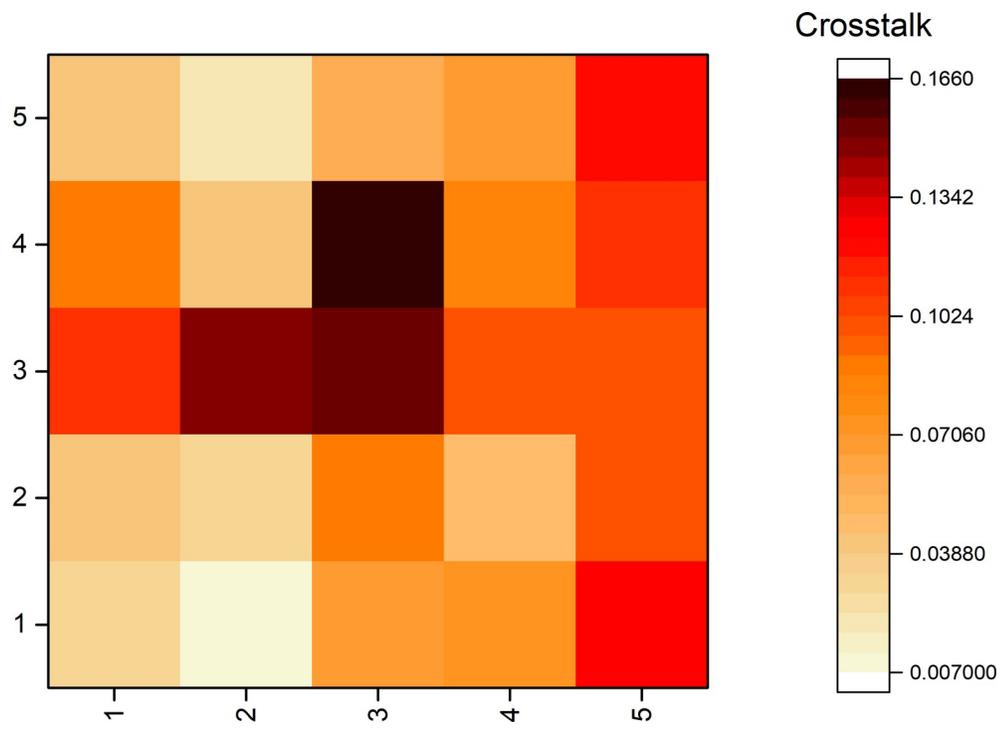

Figure 4.6: Crosstalk values calculated for all 25 pixels for an applied weight of 0.5 kg.

the appendix, the trend is followed when the pitch was in the order of "cm", also in that case the size of each pixel was much larger too. These results are interesting and should be explored in more detail to investigate any resonance effect that may be occurring. We need to perform more experiments with smaller increments in pitch to get a better understanding of the same. The novel method proposed in this study needs more rigorous analytical validation. More tests need to be performed to check how this algorithm fares for sensors with varying specifications.

*Chapter 5*

# Conclusion

As a part of this dissertation, we presented the design and fabrication of a flexible writing pad using a piezoresistive material, velostat, as the pressure sensing element. We report that characters were legible when written larger than 2 cm in size. The writing pad was found to be functional even up to a bending radius of 4 cm. This was a small implementation (5 cm $\times$ 5 cm) with only 256 pixels, however, we believe that the design principles are scalable, and a larger flexible writing pad with a similar design can be fabricated. However, this could reduce the refresh rate further. To overcome this challenge, a higher sampling frequency needs to be employed. Moreover, the spatial and temporal characteristics of the human handwriting can be exploited to recover lost data, eliminate noise and predict characters through various probabilistic models, curve-fitting techniques, and machine learning algorithms.

We have also presented a novel approach to quantify crosstalk in a sensor matrix that can be extended to any array and used across technologies. We characterised three flexible pressure sensing matrices based on a crossbar architecture using Velostat. We also used different weights to analyse the trends with changing pressure values. We observe that a pitch of 5mm gives us the least crosstalk, which is as expected. We also fully characterised the crosstalk present in a $5 \times 5$ pressure sensing matrix with a 4 mm pitch with a 0.5 kg weight. We observe the mean crosstalk value to be $0.081 \pm 0.002$, and minimum and maximum crosstalk to be 0.0071 and 0.1656. The variance of 0.002 indicates that the variation in crosstalk between pixels in the same matrix is not very large.

# Related Publications

- Mohee Datta Gupta, L. Lakshmanan, Anis Fatema, Aftab Hussain, "Flexible Writing Pad Based on a Piezoresistive Thin Film Sensor Matrix." IEEE Applied Sensing Conference (APSCON) 2023

- L. Lakshmanan, Mohee Datta Gupta, Anis Fatema, Aftab Hussain, "Characterisation and Quantification of Crosstalk on a Velostat-based Flexible Pressure Sensing Matrix." IEEE International Conference on Flexible and Printable Sensors and Systems (FLEPS). 2023

- Anis Fatema, Mohee Datta Gupta, Shirley Chauhan, Aftab Hussain, "Investigation of Reliability of a Velostat-based Flexible Pressure Sensor Mat." IEEE Transactions on device and material reliability 2023 (in review)

# Other Publications

- Datta Gupta, M., Mishra, R.B., Kuriakose, I. and Hussain, A.M., 2022. "Determination of thermal and mechanical properties of SU-8 using electrothermal actuators." MRS Advances, 7(28), pp.591-595.

- Anis Fatema, Saurabh B. Mishra, Mohee Datta Gupta, Aftab Hussain, "Polypyrrole-based Cotton Flexible Pressure Sensor using In-situ Chemical Oxidative Polymerization" IEEE International Conference on Flexible and Printable Sensors and Systems (FLEPS). 2023

*Appendix A*

# Additional Results for Crosstalk

## A.1 Introduction

The experiments discussed in this section were conducted as a part of another project in our research group where we investigated the reliability of a velostat-based flexible pressure sensor array to assess its suitability for various applications involving reliability as an essential criterion. The performance was examined by testing the sensor's response for one year. The reliability of the sensor determines its quality and durability. It is an essential and critical factor to be considered when designing any sensor. The details of the findings of this project are beyond the scope of this thesis. In this section, we are going to look at the crosstalk values that were calculated on the data gathered from the experiments using the algorithm proposed in Chapter 4 of this thesis. This is an attempt to demonstrate with an example how the algorithm can be generalized and extended to any kind of pressure sensor with a crossbar architecture.

## A.2 Experiments

To collect the crosstalk data, we designed a 3 × 3 sensor matrix with each pixel of size 10 mm × 10 mm. Therefore, each pixel had a sensing area of 100 mm$^2$. Five such sensors were created with pitch varying from 1cm, 2cm, 3cm, 4cm, and 5cm. A 10 mm × 10 mm acrylic block with a height of 2 mm, was cut so that an individual sensor pixel could be activated accurately. Moreover this enables us to apply a specific weight which was evenly distributed on a particular pixel, and hence standardize the force on each pixel for all experiments. The acrylic block should be placed exactly on the sensing area to get accurate readings as a slight shift in the block may result in garbage readings. We placed a weight of 0.5 kg on the acrylic block to collect the readings to calculate the crosstalk. For each mat, there are 9 pixels.

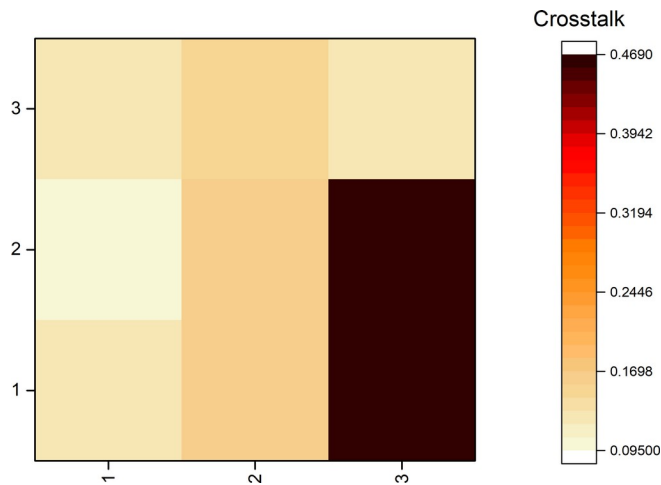

Figure A.1: Crosstalk values calculated for all 9 pixels for a 3x3 sensor matrix with pitch of 1 cm and load of 0.5kg.

The weight was placed one-by-one, using the acrylic block and the crosstalk was calculated for each of the 9 pixels, for all the 5 mats with varying pitch.

## A.3   Results and Discussion

In Fig A.1- A.5, the crosstalk values for all 9 pixels are displayed for pitch lengths of 1cm, 2 cm, 3 cm, 4 cm, and 5 cm respectively. As expected from Chapter 4, the crosstalk decreases as the pitch length increases which is illustrated in Fig A.6, where the mean crosstalk is plotted against varying pitch length of the sensor matrix. The actual data for the same is provided in Table A.1. We can observe that the crosstalk practically vanishes from a pitch of 5 cm. Therefore for the reliability experiments all the sensor mats were designed with a pitch of 5 cm to minimize crosstalk.

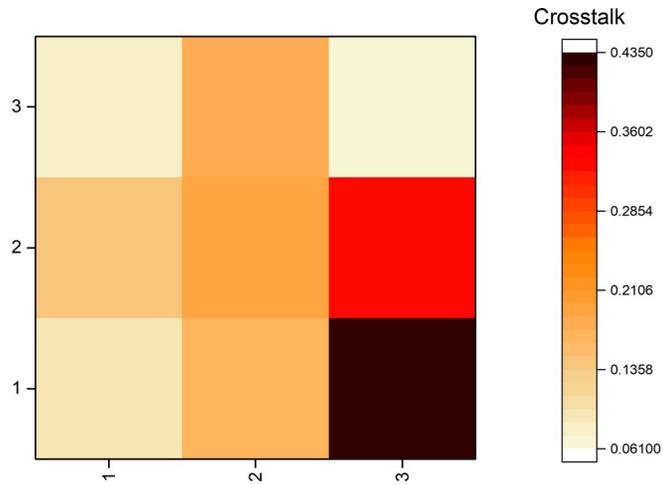

Figure A.2: Crosstalk values calculated for all 9 pixels for a 3x3 sensor matrix with pitch of 2 cm and load of 0.5kg.

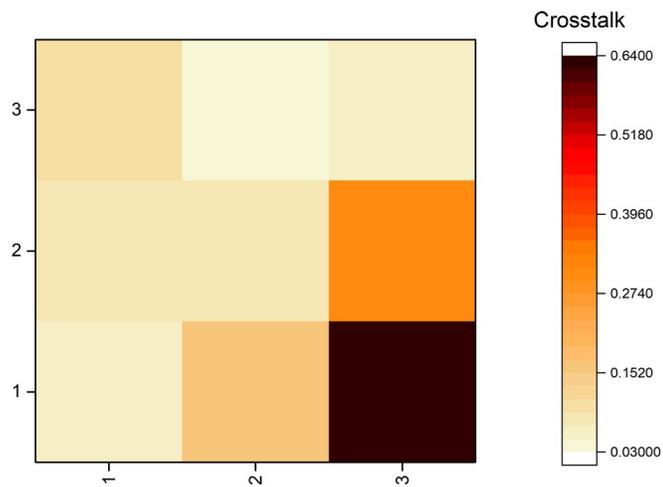

Figure A.3: Crosstalk values calculated for all 9 pixels for a 3x3 sensor matrix with pitch of 3 cm and load of 0.5kg.

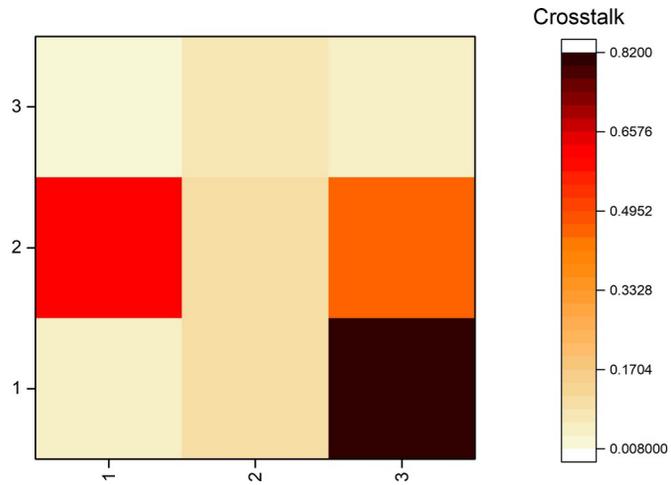

Figure A.4: Crosstalk values calculated for all 9 pixels for a 3x3 sensor matrix with pitch of 4 cm and load of 0.5kg.

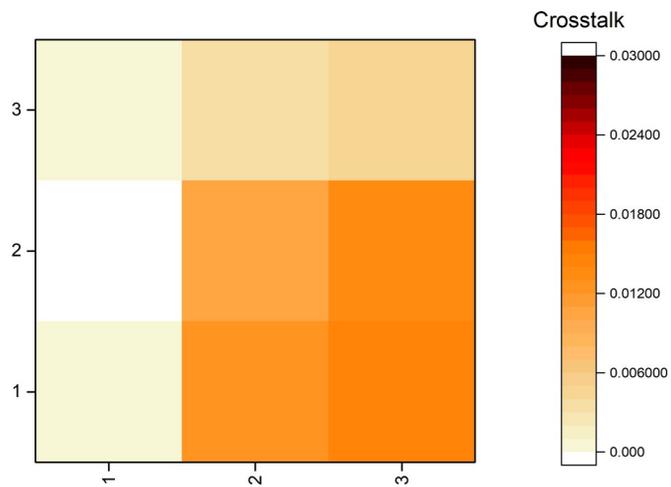

Figure A.5: Crosstalk values calculated for all 9 pixels for a 3x3 sensor matrix with pitch of 5 cm and load of 0.5kg.

Table A.1: Mean Crosstalk for varying pitch of sensor mat

| Pitch Length (in cm) | Mean Crosstalk |
|---|---|
| 1 | 0.2075±0.0192 |
| 2 | 0.1860±0.0133 |
| 3 | 0.1709±0.0329 |
| 4 | 0.1932±0.0635 |
| 5 | 0.0099±0.0008 |

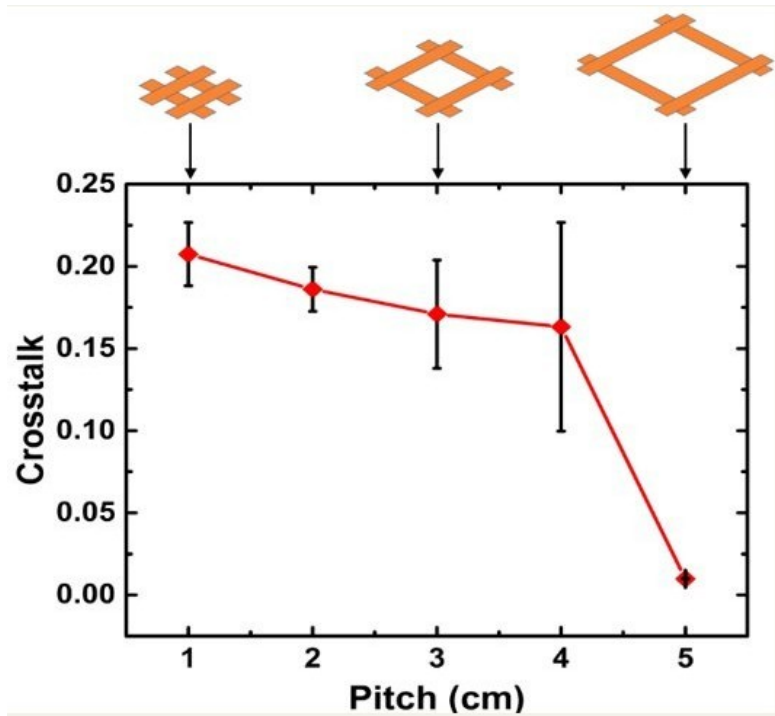

Figure A.6: Mean crosstalk values calculated for sensor matrices with varying line pitches